\documentclass[twocolumn,twocolappendix]{aastex631}

\submitjournal{ApJ}
\usepackage{cuted}
\usepackage{amsmath}
\usepackage{booktabs}
\usepackage{float}
\usepackage{longtable}
\usepackage{cuted}

\begin{document}


\title{The Impact of Plasma Angular Momentum on Magnetically Arrested Flows and Relativistic Jets in Hot Accretion Flows Around Black Holes}

\author[0000-0003-2776-082X]{Ho-Sang Chan}
\email{hschanastrophy1997@gmail.com}
\altaffiliation{Croucher Scholar}
\affiliation{JILA, University of Colorado and National Institute of Standards and Technology, 440 UCB, Boulder, CO 80309-0440, USA}
\affiliation{Department of Astrophysical and Planetary Sciences, University of Colorado, 391 UCB, Boulder, CO 80309-0391, USA}

\author[0000-0001-9446-4663]{Prasun Dhang}
\affiliation{JILA, University of Colorado and National Institute of Standards and Technology, 440 UCB, Boulder, CO 80309-0440, USA}
\affiliation{Department of Astrophysical and Planetary Sciences, University of Colorado, 391 UCB, Boulder, CO 80309-0391, USA}

\author[0000-0003-3903-0373]{Jason Dexter}
\affiliation{JILA, University of Colorado and National Institute of Standards and Technology, 440 UCB, Boulder, CO 80309-0440, USA}
\affiliation{Department of Astrophysical and Planetary Sciences, University of Colorado, 391 UCB, Boulder, CO 80309-0391, USA}

\author[0000-0003-0936-8488]{Mitchell C. Begelman}
\affiliation{JILA, University of Colorado and National Institute of Standards and Technology, 440 UCB, Boulder, CO 80309-0440, USA}
\affiliation{Department of Astrophysical and Planetary Sciences, University of Colorado, 391 UCB, Boulder, CO 80309-0391, USA}


\begin{abstract}


In certain scenarios, the accreted angular momentum of plasma onto a black hole could be low; however, how the accretion dynamics depend on the angular momentum content of the plasma is still not fully understood. We present three-dimensional, general relativistic magnetohydrodynamic simulations of low angular momentum accretion flows around rapidly spinning black holes (with spin $a = +0.9$). The initial condition is a Fishbone-Moncrief (FM) torus threaded by a large amount of poloidal magnetic flux, where the angular velocity is a fraction $f$ of the standard value. For $f = 0$, the accretion flow becomes magnetically arrested and launches relativistic jets but only for a very short duration. After that, free-falling plasma breaks through the magnetic barrier, loading the jet with mass and destroying the jet-disk structure. Meanwhile, magnetic flux is lost via giant, asymmetrical magnetic bubbles that float away from the black hole. The accretion then exits the magnetically arrested state. For $f = 0.1$, the dimensionless magnetic flux threading the black hole oscillates quasi-periodically. The jet-disk structure shows concurrent revival and destruction while the gas outflow efficiency at the event horizon changes accordingly. For $f \geq 0.3$, we find that the dynamical behavior of the system starts to approach that of a standard accreting FM torus. Our results thus suggest that the accreted angular momentum is an important parameter that governs the maintenance of a magnetically arrested flow and launching of relativistic jets around black holes.


\end{abstract}



\section{Introduction} \label{sec:intro}


The difference between radio-loud and radio-quiet galaxies lies in the ability of the former to produce large-scale radio jets, the energy of which is a significant fraction of the total bolometric luminosity \citep{1995ApJ...438...62W}. It is now widely accepted that relativistic jets are produced via the Blandford–Znajek process \citep{1977MNRAS.179..433B}, in which poloidal magnetic fields threading the black hole extract angular momentum and spin energy from the hole and help accelerate plasma to relativistic speeds. More importantly, numerical simulations of black hole accretion have found that the accumulation of intense magnetic flux on the black hole can form a magnetically arrested disk (MAD) state \citep{tchekhovskoy2011efficient}. This provides an ideal scenario for the Blandford-Znajek process to operate and launch bipolar collimated outflows with a jet efficiency greater than $100$\,$\%$. The production of relativistic jets thus seems to rely on two critical parameters: 1) the spin of the black hole and 2) the amount of magnetic flux threading its event horizon.

The Event Horizon Telescope Collaboration (EHTC) released spatially resolved, polarized images of the supermassive black hole at our (radio-quiet) Galactic Center, Sagittarius A* (Sgr~A*). These results suggest that the black hole could be spinning \citep{2022ApJ...930L..16E}, and the magnetic field close to the event horizon is dynamically important \citep{akiyama2024first}. Thus, the accreting plasma should be strongly magnetized so that the black hole should have entered the MAD state and launched relativistic jets. However, no radio jet has been observed to emanate from Sgr~A*. The fact that Sgr~A* contains dynamically important magnetic fields but no relativistic jet is observed seems contradictory. In view of this, we are motivated to examine whether some other factor is preventing the formation of a relativistic jet. One such possibility is that the black hole is spinning slowly \citep{narayan2022jets}; An alternative, which we focus on here, is that the accreting gas itself lacks angular momentum.


Previous studies have shown that accreted angular momentum plays an important role in the subsequent dynamical evolution of the accreting black hole, especially when magnetic effects are included. \citet{igumenshchev2002three} examined radiatively inefficient spherical accretion onto a black hole, where the plasma is immersed in a weak vertical magnetic field. They found that the presence of a magnetic field significantly alters the accretion dynamics compared to the Bondi solution. Magnetic fields are amplified through flux freezing, reaching equipartition with the gas pressure and initiating reconnection. The reconnection is self-sustained via gas convection and further magnetic reconnection. The resulting dynamics exhibits a highly subsonic inflow velocity and a mass accretion rate well below the value predicted by the Bondi solution. Later, \citet{proga2003accretionMHD} studied spherical accretion onto black holes with a modest amount of angular momentum. They showed that threading the initially isotropic and homogeneous plasma with a weak poloidal magnetic field drastically reduces mass accretion rates. The spherical free-fall of the material drags the magnetic field to the horizon, forming a magnetic barrier that prohibits matter from falling onto the black hole. Bipolar outflows are launched, pushing away free-falling material along the pole. The magnetorotational instability redistributes the angular momentum of material with too much angular momentum to be accreted, forming an accretion disk around the black hole. Notably, \citet{2003ApJ...596L.207P, 2011MNRAS.415.1228P} discovered a class of magnetically frustrated convection, in which magnetic shear stresses counteract buoyant motions. This leads to an inflow of energy despite the presence of convective motion. Neither steady outflow nor rotationally supported flow has been observed.

The accretion of low angular momentum plasma onto \textit{spinning} black holes, where the frame-dragging effects are self-consistently taken into account, has only recently been examined\footnote{See also \citet{Kwan_2023}}. \citet{ressler2021magnetically} reported a series of GRMHD simulations of spherical accretion onto spinning black holes, with the large-scale coherent magnetic field making different angles $\psi$ with the black hole's spin axis. They found that magnetic fields are amplified via flux freezing, reconnect, and drive turbulence. Relativistic jets propagate to a few hundred gravitational radii ($GM/c^2$) before dissipating due to kink instability. Except for the `sweet spot' angle $\psi = 60^\circ$, none of their models sustain a MAD state, but are otherwise semi-MAD ($\phi_{\rm BH} \gtrsim 20$) with a gas outflow efficiency as low as $10$\,\%. For $\psi \sim 30^\circ$ or $90^\circ$, an initial bump followed by a significant drop in $\Phi_{\rm BH}$ is observed. In several instances, the infalling gas fills one side of the bipolar jet, but the polar region is quickly evacuated, creating a semi-sustainable outflow. Meanwhile, \citet{2023arXiv231011487L} presented a high-resolution GRMHD simulation of spherical accretion threaded by a weak vertical magnetic field. The initially launched relativistic jet propagates to $\sim 1500$\,$GM/c^2$. The accretion has a saturated $\phi_{\rm BH} \sim 50$ for a long duration ($\sim 34000$\,$GM/c^3$). After that, the power of the jet shuts off, with an associated drop in $\phi_{\rm BH}$ to a level of $\sim 20$ and $\eta \sim 10-15$\,\%. In the aftermath, the transient jets are weak and dissipate within the Bondi radius. The accretion attains the MAD state, but only for a finite duration. These results suggest that the angular momentum content is a crucial parameter, in addition to the magnetic flux and the black hole spin, that governs strongly magnetized accretion and the launching of relativistic jets. However, why the angular momentum content is important and what distinguishes between accretion with low and high angular momentum content is still not fully understood.


Additionally, these previous studies assumed an initial condition of a homogeneous, spherically symmetric plasma with either no or a non-zero but small amount of angular momentum that depends on the polar angle \citep{proga2003accretionMHD}. However, to our knowledge, the accretion of low angular momentum and strongly magnetized plasma with a torus as the initial condition. has not been thoroughly examined. Our motivation is to provide the black hole with a well-defined mass reservoir, but with the angular momentum content being variable as a free parameter. From a technical point of view, although simulations spanning orders of magnitude in spatial scales that connect the black hole horizon to the Bondi radius are possible, they often require considerable computational resources. Therefore, it seems more economical to study the accretion of low angular momentum plasma onto black holes via the torus setup, thus enabling us to explore a wider parameter space.

Our current study focuses on the accretion of low angular momentum and strongly magnetized plasma onto spinning black holes. Specifically, we are interested in examining how the accreted angular momentum content of the plasma affects the 1) accretion dynamics, 2) maintenance of the MAD state, and 3)  launching of relativistic jets.

To achieve this, we present a series of GRMHD simulations of plasma accretion onto spinning black holes with varying angular momentum content to systematically study how angular momentum affects the resulting accretion flow. As we will show, the angular momentum content plays an important role in governing magnetic flux saturation and recycling, thus greatly impacting the maintenance of the MAD state and the launching of bipolar jets. The paper is structured as follows: Section \ref{sec:method} presents the tools, simulation setup, and basic diagnostic quantities. In Section \ref{sec:results}, we discuss the overall diagnostics of our simulation results, including time-series analysis and flow morphology. We will further analyze two interesting cases and show how they differ from models with more angular momentum. One model becomes magnetically arrested and produces powerful jets for a very short duration before the jet is destroyed; the other model shows quasi-periodic transitions between the magnetically arrested state and non-arrested state, leading to jet disruption and revival accordingly. We discuss the physical implications in Section \ref{sec:physical} and summarize our results in Section \ref{sec:conclude}. 

During the preparation of this manuscript, we found a similar work submitted on arXiv \citep{galishnikova2024stronglymagnetizedaccretionlow}. We remark that although both studies are interested in the effect of angular momentum on strongly magnetized flow, we approach this problem with different initial setups and performed different sets of analyses. Therefore, we believe that our manuscript should serve as a complementary work to their study.
 

\section{Methodology} \label{sec:method}


\subsection{Simulation Setup} \label{subsec:setup}

\begin{deluxetable}{ccccc}[htb!]
\tablecaption{List of simulations and their model parameters, including the ratio $f$ of the initial angular velocity to the standard value, the volume-averaged specific angular momentum $l_{\rm avg}$, the specific angular momentum measured at the density maximum $l_{0}$, and the circularization radius $r_{c}$ measured with respect to $l_{0}$. For $f = 0.0 - 0.3$, there is no solution for $r_{c}$. All models have a black hole spin $a = +0.9$. \label{tab:models}}
\tablewidth{0pt}
\tablehead{
\colhead{Model} & \colhead{$f$} & \colhead{$l_{\rm avg}$} & \colhead{$l_{0}$} & \colhead{$r_{c}$}
}
\startdata
a$09$f$00$ & $0.0$ & $-0.008$ & $-0.112$ & - \\
a$09$f$01$ & $0.1$ & $0.484$ & $0.376$ & - \\ 
a$09$f$03$ & $0.3$ & $1.468$ & $1.349$ & - \\ 
a$09$f$05$ & $0.5$ & $2.452$ & $2.311$ & $3.599$ \\ 
a$09$f$07$ & $0.7$ & $3.435$ & $3.254$ & $8.682$ \\ 
a$09$f$10$ & $1.0$ & $4.909$ & $4.498$ & $18.1$
\enddata
\end{deluxetable}


To model the accretion of plasma onto spinning black holes, we solve the ideal GRMHD equations (with $G$, $c$, and the black hole mass $M_{\rm BH} = 1$) using the open-source code \texttt{Athena++} \citep{stone2020athena++}:
\begin{equation}
\begin{aligned}
    \partial_{t}(\sqrt{-g}\rho u^{t}) &= -\partial_{i}(\sqrt{-g}\rho u^{i}), \\
    \partial_{t}(\sqrt{-g}T^{t}_{\;\;\nu}) &= -\partial_{i}(\sqrt{-g} T^{i}_{\;\;\nu}) + \frac{1}{2}\sqrt{-g}T^{\kappa\lambda}\partial_{\nu}g_{\kappa\lambda}, \\
    \partial_{t}(\sqrt{-g}B^{i}) &= -\partial_{j}[\sqrt{-g}(b^{j}u^{i} - b^{i}u^{j})], \\
    \partial_{i}(\sqrt{-g}B^{i}) &= 0.
\end{aligned}
\end{equation}
in which $\sqrt{-g}$ is the metric's determinant, $g_{\mu\nu}$ is the metric tensor, $\rho$ is the rest-mass density, $B^{i}$ is the magnetic field in the coordinate frame, $u^{\mu}$ is the $4-$velocity, and $b^{\nu}$ is the magnetic $4-$vector, which relates to the magnetic $3-$vector as
\begin{equation}
\begin{aligned}
    b^{t} = g_{i\mu}B^{i}u^{\mu}, \\
    b^{i} = \frac{B^{i} + b^{t}u^{i}}{u^{t}}.
\end{aligned}
\end{equation}
We adopt a spherical-polar version of the Kerr-Schild coordinates $(t, r, \theta, \phi)$. The stress-energy tensor is
\begin{equation}
    T^{\mu\nu} = (\rho h + b^{2})u^{\mu}u^{\nu} + (P + \frac{b^{2}}{2})g^{\mu\nu} - b^{\mu}b^{\nu},
\end{equation}
where $h = 1 + \gamma/(\gamma - 1)P/\rho$ is the specific enthalpy, $P = \rho\epsilon (\gamma - 1)$ is the gas pressure, $\epsilon$ is the specific internal energy, $b^{2} = b^{\mu}b_{\mu}$, and $g^{\mu\nu}$ is the inverse metric. 

The initial conditions for the simulations are the standard Fishbone-Moncrief (FM) torus \citep{fishbone1976relativistic}. To vary the angular momentum content of the torus, we set $u^{\phi}$ as a fraction $f$ of the standard value, spanning $f = 0.0, 0.1, 0.3, 0.5, 0.7, 1.0$. The torus has an inner radius of $8.0$\,$GM/c^{2}$ the pressure maximum is at $18.1$\,$GM/c^{2}$, while the outer radius is at around $650$\,$GM/c^{2}$, all measured from the center of the black hole along the mid-plane. We define the effective Bondi radius of the torus as $R_{B} = 2GM/c_{s}^{2}$ where $c_{s}$ is the sound speed measured at the radius of maximum pressure. We find $R_{B} \sim 127$\,$GM/c^{2}$, while the ratio of  the radius of pressure maximum to $R_{B}$ is roughly $0.142$. Note that we also calculate $R_{B}$ based on the floor value of $\rho$ and $P$, and it gives $R_{B} \sim 122$\,$GM/c^{2}$. The torus is threaded with a single, large-scale magnetic flux loop defined by a vector potential \citep{2019ApJ...874..168W}:
\begin{equation}
\begin{aligned}
    A_{\phi} &= C_{B}\text{max}(P - P_{\text{cut}}, 0)^{q_{p}}r^{q_{r}} \\
    &\ \quad \sin{\theta}\sin{\theta_{r}}\sin{\theta_{\rm th}}, \\
    \theta_{r} &= \pi n_{r} L(r; r_{B,\text{min}}, r_{B,\text{max}}), \\
    \theta_{\text{th}} &= \pi n_{\theta} L(\theta; \theta_{B,\text{min}}, \pi - \theta_{B,\text{min}}), \\
    L(q; q_{\text{min}}, q_{\text{max}}) &= \text{min}\left[\text{max}\left(\frac{q - q_{\text{min}}}{q_{\text{max}} - q_{\text{min}}}\right), 1\right],
\end{aligned}
\end{equation}
where we set $P_{\text{cut}} = 10^{-8}$, $q_p = 0.5$, $q_r = 2.0$, $n_r = n_\theta = 1$, $r_{B,\text{min}} = 8.5$, $r_{B,\text{max}} = 2000$, $\theta_{B,\text{min}} = 0.6$, and $C_B = 0.5$. Using these numbers, the average plasma $\beta$, which is defined as the maximum gas to maximum magnetic pressure, is about $1330$. The volume averaged plasma $\beta$, which is defined as $2\int PdV/\int b^{2}dV$, with $dV = \sqrt{-g}drd\theta d\phi$, is about $30$. We employ a radius-dependent floor on the mass density and pressure such that
\begin{equation}
\begin{aligned}
    P_{\rm floor} &= \text{max}(10^{-6}r^{-2.5}, 10^{-10}), \\
    \rho_{\rm floor} &= \text{max}(10^{-4}r^{-1.5}, 10^{-8}),
\end{aligned}
\end{equation}
and we also impose limits on the pressure and mass density based on the plasma $\beta = 2P/b^{2}$, plasma $\sigma = b^{2}/\rho$, and the Lorentz factor $\Gamma$. In particular, we set a ceiling of $\sigma \leq 100$, $\Gamma \leq 50$, and a floor of $\beta \geq 0.001$. 

We use the 3rd order Piecewise Parabolic Method to interpolate primitive variables to cell interfaces and the HLLE Riemann solver to compute fluxes at cell interfaces. To march the system of equations in time, we use the 2nd Order Van Leer time-integrator with a Courant–Friedrichs–Lewy number of $0.3$. The simulations are evolved for a time interval of $\tau = 20000$\,$GM/c^{3}$.

\begin{figure}[htb!]
    \centering
    \includegraphics[width=1.0\linewidth]{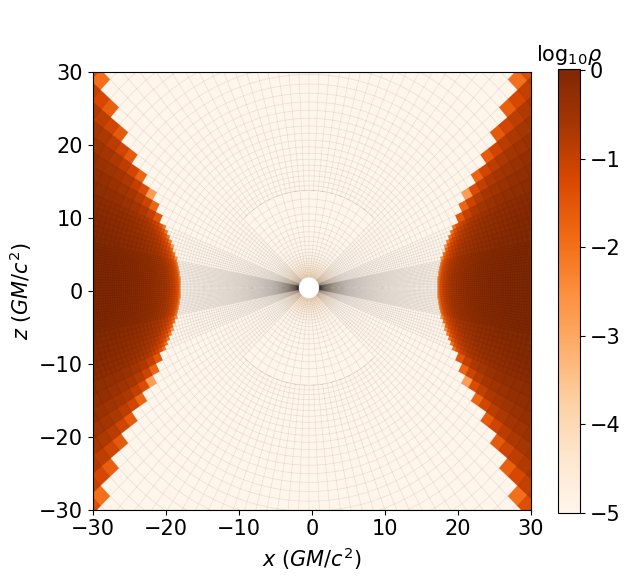}
    \caption{The initial density distribution of the torus (in the log$_{10}$ scale) and grid setup (shown as transparent lines) along the $x-z$ plane. \label{fig:grid}}
\end{figure}

The inner boundary along the $r$ coordinate, $r_{\rm in}$, is placed within the outer horizon $r_{\rm BH}$ of the black hole, where $r_{\rm BH} = 1 + \sqrt{1 - a^{2}}$. The outer boundary is located at $1000$\,$GM/c^{2}$. We cover the full polar and azimuthal angles. Polar boundary conditions are applied along the pole, while periodic boundary conditions are applied along the azimuthal direction. The inner and outer radial boundaries are set to allow material to flow out of the boundary but not vice versa. Finally, our simulation uses three layers of static mesh refinement. The base resolution is $72\times32\times32$, while the effective resolution is $576\times256\times256$. All levels of refinement cover the full azimuthal angle. We show the $x-z$ slice of the initial condition with grid structures appended in Figure \ref{fig:grid} for reference. The parameters of the simulations are shown in Table \ref{tab:models}.

%


\subsection{Basic Diagnostics} \label{subsec:diag}

We outline several diagnostic quantities of interest. First, we define the plasma fluxes \citep{narayan2012grmhd, dhang2023magnetic} of mass ($\dot{M}$), magnetic field ($\Phi_{\rm BH}$), and energy ($\dot{E}$):
\begin{equation} \label{eqn:diag}
\begin{aligned} 
    \dot{M} &= -\int_{0}^{2\pi}\int_{0}^{\pi}\rho u^{r}\sqrt{-g}d\theta d\phi, \\ 
    \Phi_{\rm BH} &= \sqrt{4\pi}\int_{0}^{2\pi}\int_{0}^{\pi}|B^{r}|\sqrt{-g}d\theta d\phi, \\
    \dot{E} &= \int_{0}^{2\pi}\int_{0}^{\pi}T^{r}_{t}\sqrt{-g}d\theta d\phi;
\end{aligned}
\end{equation}
these fluxes are all evaluated at $r_{\rm BH}$. Also, note that we have adopted the sign convention in \citet{dhang2023magnetic} so that the accreted specific energy $e = \dot{E}/\dot{M}$. Additionally, the dimensionless magnetic flux threading the event horizon is defined as $\phi_{\rm BH} = \Phi_{\rm BH}/(2\sqrt{\langle \dot{M}\rangle})$, and the horizon gas outflow efficiency is given as $\eta = (\dot{M} - \dot{E})/\langle \dot{M}\rangle$ \citep{tchekhovskoy2011efficient}. Here, brackets refer to time-averaging, and the averaging for $\dot{M}$ is done with a window of $500$ $GM/c^{3}$. We will study the one-dimensional profile of an arbitrary quantity $Q$ by performing the density-weighted  shell average:
\begin{equation}
    \langle Q\rangle_{\theta, \phi}(r) = \frac{\int\rho Q\sqrt{-g}d\theta d\phi}{\int\rho \sqrt{-g}d\theta d\phi}.
\end{equation}
To study the two-dimensional flow structure, we compute the azimuthal average of an arbitrary quantity $Q$ as
\begin{equation}
    \langle Q\rangle_{\phi}(r, \theta) = \frac{\int_{0}^{2\pi}Qd\phi}{\int_{0}^{2\pi}d\phi} .
\end{equation}
We also visualize the velocity and magnetic field lines. The coordinate velocity $V^{i}$, where $i = r, \theta, \phi$, is given by $V^{i} = u^{i}/u^{t}$. The $3-$magnetic field is obtained from the GRMHD equations. These vectors are given in the coordinate basis. To visualize $V^{i}$ and $B^{i}$ within a quasi-orthonormal basis, we scale the $r, \theta,$ and $\phi$ components of the three velocities and magnetic fields so that \citep{dhang2023magnetic}
\begin{equation} \label{eqn:vector}
\begin{aligned}
    V_{r} = \sqrt{g_{rr}}V^{r}, V_{\theta} &= \sqrt{g_{\theta\theta}}V^{\theta}, V_{\phi} = \sqrt{g_{\phi\phi}}V^{\phi} \\
    B_{r} = \sqrt{g_{rr}}B^{r}, B_{\theta} &= \sqrt{g_{\theta\theta}}B^{\theta}, B_{\phi} = \sqrt{g_{\phi\phi}}B^{\phi}.
\end{aligned}
\end{equation}
%



\section{Results} \label{sec:results}


\subsection{Time-series Analysis} \label{sec:timeseries}
\begin{figure*}[htb!]
    \centering
    \gridline{
    \fig{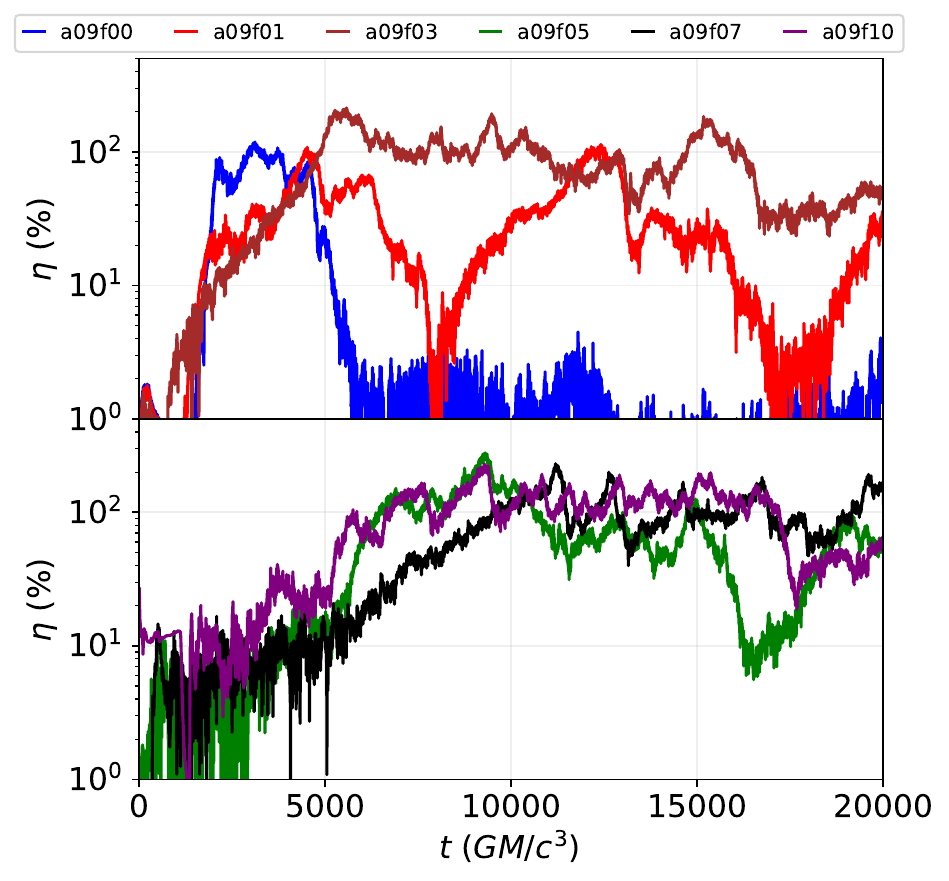}{0.5\textwidth}{(a)}
    \fig{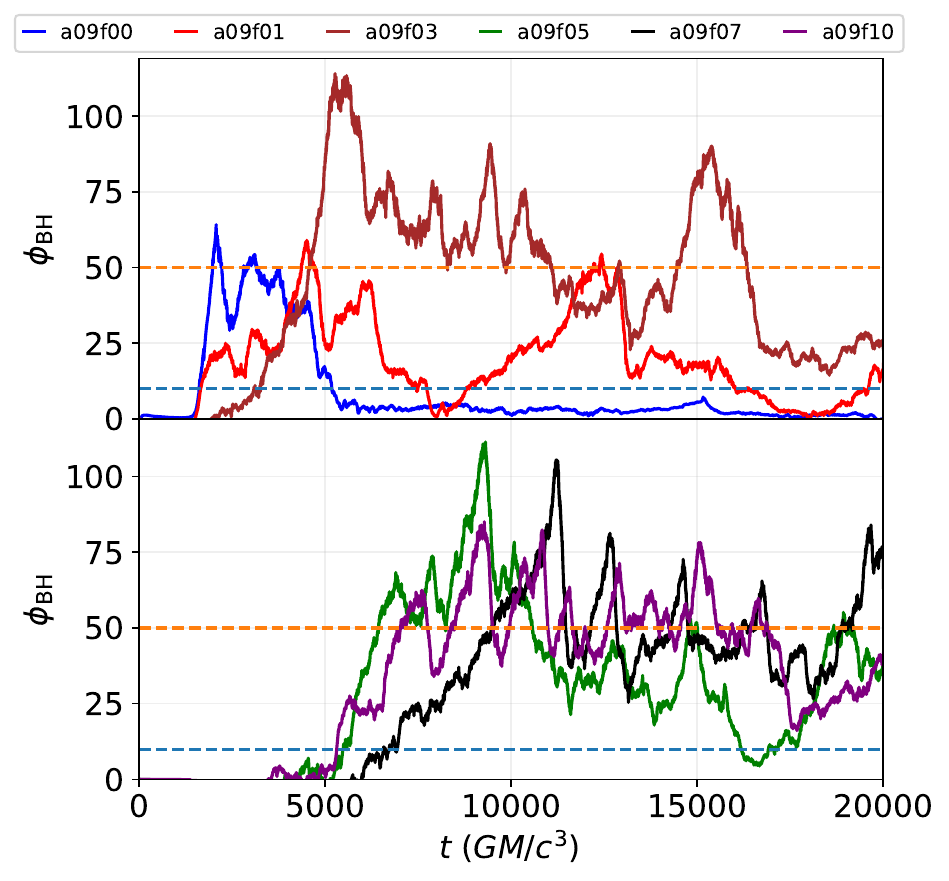}{0.5\textwidth}{(b)}}
    \caption{(a) The horizon gas outflow efficiency $\eta$ (in $\%$) and (b) the dimensionless magnetic flux threading the event horizon $\phi_{\rm BH}$. In (b), the blue (orange) dashed-dotted horizontal lines represent $\phi_{\rm BH} = 10$ $(50)$. For model $a$09$f$00, $\eta$ and $\phi_{\rm BH}$ drop significantly at $\sim 5000$\,$GM/c^{3}$; for model $a$09$f$01, $\eta$ and $\phi_{\rm BH}$ oscillate quasi-periodically. \label{fig:timeseries}}
\end{figure*}
\begin{figure}[htb!]
    \centering
    \includegraphics[width=1.05\linewidth]{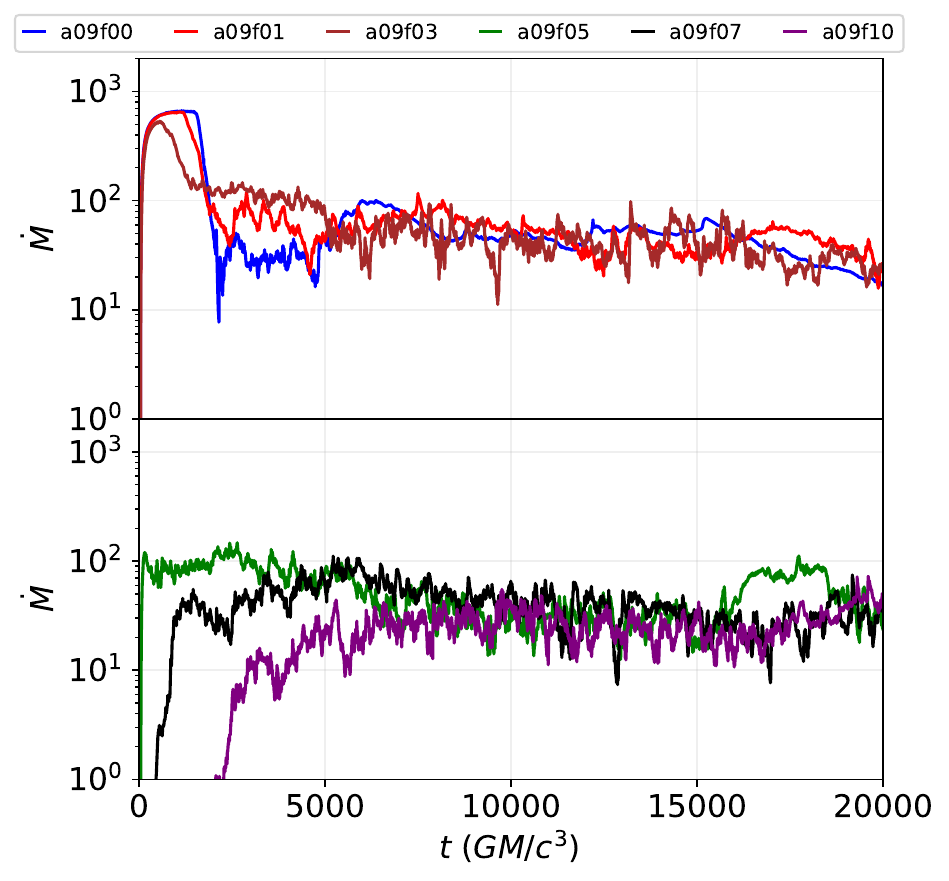}
    \caption{Same as Figure \ref{fig:timeseries}, but for $\dot{M}$. Low angular momentum models (e.g., $a$09$f$00 and $a$09$f$01) show a significant $\dot{M}$ at the beginning, then it drops substantially. High angular momentum models (e.g., $a$09$f$07 and $a$09$f$10) show a gradually building but smaller $\dot{M}$. \label{fig:timeseries2}}
\end{figure}

We show the time-series diagnostics in Figure \ref{fig:timeseries}. We find that low angular momentum models (e.g., $a$09$f$00 and $a$09$f$01) generate powerful jets, reaching $\eta \sim 100$\% at the beginning of the simulation. For models with more angular momentum content (e.g., $a$09$f$07 and $a$09$f$10), $\eta$ increases more slowly and reaches approximately $100$\,\% at a later stage. The temporal evolution of $\eta$ is consistent with that of $\phi_{\rm BH}$. When $\eta$ is about $100$\,\%, we find $\phi_{\rm BH}$ reaching the saturated value of $\sim 50$. This confirms that horizon-penetrating magnetic fields are responsible for extracting energy from the rotating black hole and launching relativistic jets.

For models with low angular momentum content, the earlier time in reaching $\eta \sim 100$\% is due to the lack of centrifugal barriers that prohibit plasma from freely falling onto the black hole. The prompt infalling of plasma thus carries a considerable amount of poloidal magnetic flux with it, filling the black hole within a short period of time. This could also be seen from the time series of $\dot{M}$ (see Figure \ref{fig:timeseries2}) and $\phi_{\rm BH}$. We find that low angular momentum models have a large $\dot{M}$ at the beginning of the simulation. At the same instant, $\phi_{\rm BH}$ is building up quickly, hinting at a large amount of magnetic flux threading the event horizon. After that, $\dot{M}$ drops significantly, indicating the development of a magnetically arrested flow. In contrast, the rise time of $\phi_{\rm BH}$ for models with more angular momentum content is longer, and they have smaller initial, gradually increasing $\dot{M}$.

However, we find that powerful relativistic jets for models $a$09$f$00 and $a$09$f$01 are not sustained, and they do not attain the MAD state. The gas outflow efficiency $\eta$ for model $a$09$f$00 drops substantially at $t \sim 5000\,GM/c^{3}$ to a level of $\sim 1$\%. The gas outflow efficiency for model $a$09$f$01 shows a quasi-periodic cycle, oscillating between $\sim 1$\% and $\sim 100$\%. Both models show a reduction of $\phi_{\rm BH}$ from their saturated value to $\lesssim 10$. Only models with a large enough angular momentum content (e.g., $a$09$f$07 - $a$09$f$10) can sustain powerful relativistic jets and attain the MAD state. We find that $\dot{M}$ for models $a$09$f$00 and $a$09$f$01 does not change appreciably when $\eta$ and $\phi_{\rm BH}$ drop. The disruption of the magnetically arrested flow and the relativistic jet for models $a$09$f$00 and $a$09$f$01 is due to the lack of poloidal flux penetrating the black hole. We remark that models $a$09$f$03 and $a$09$f$05 seem to be in the transition region between low and high angular momentum flow.


\subsection{Flow Morphology} \label{sec:maps}
\begin{figure*}[htb!]
    \centering
    \includegraphics[width=1.0\linewidth]{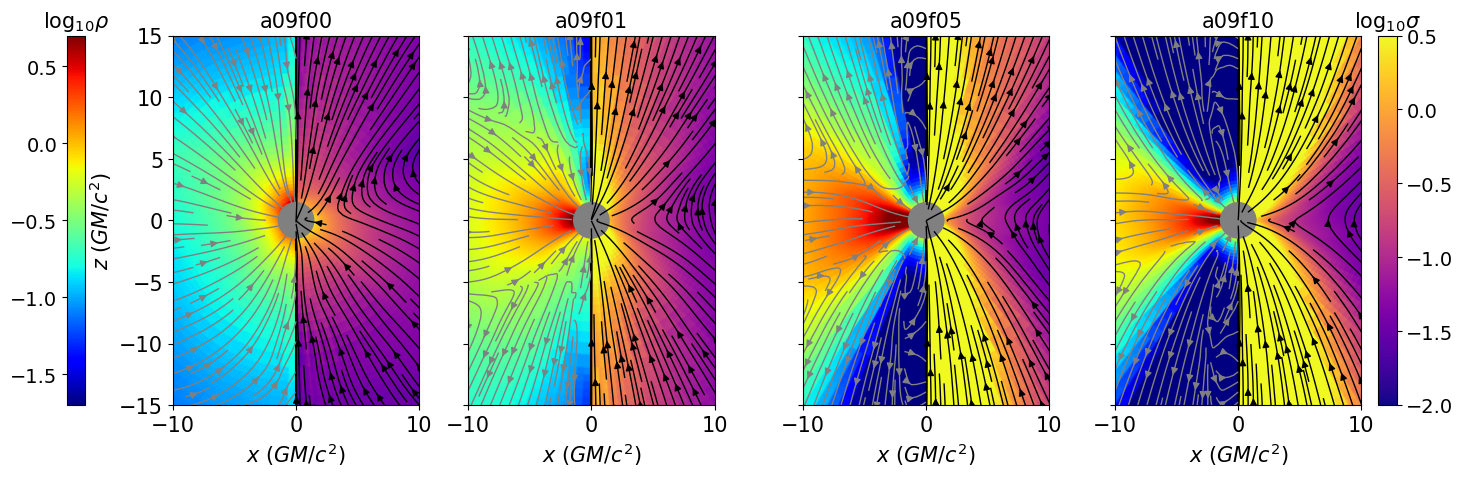}
    \caption{The $\phi$-averaged $\rho$ and $\sigma$ contours (log$_{10}$ scale) appended with velocity and magnetic field lines, respectively, for models $a$09$f$00, $a$09$f$01, $a$09$f$05, and $a$09$f$10. In each subplot, the left (right) panel shows $\rho$ ($\sigma$). All data are further time-averaged between $15000$\,$GM/c^{3}$ and $20000$\,$GM/c^{3}$, with a cadence of $10$\,$GM/c^{3}$. Models are labeled at the top of each subplot. Model $a$09$f$00 does not have a jet-disk structure, while model $a$09$f$01 has a weak bipolar outflow with a small opening angle. Only models with enough angular momentum content form a jet-disk structure with a strong bipolar outflow.\label{fig:rho-phi}}
\end{figure*}
\begin{figure}[htb!]
    \centering
    \includegraphics[width=1.0\linewidth]{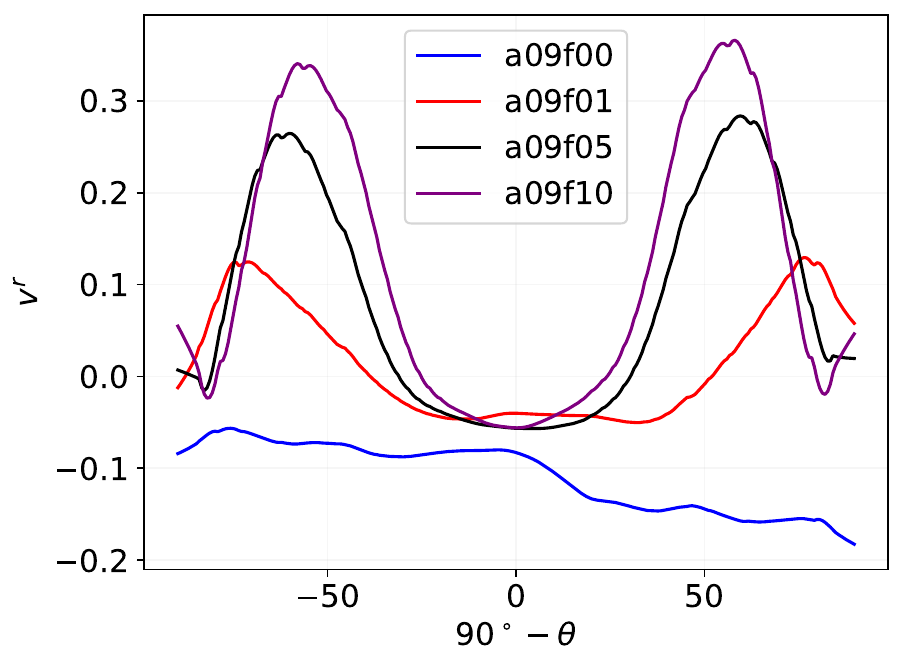}
    \caption{The $\phi$- and time-averaged radial velocity $v^{r} = u^{r}/u^{t}$ against polar angle $\theta$ measured at $r = 10$\,$GM/c^{2}$. The time averaging is done between $15000 - 20000$\,$GM/c^{3}$ with a cadence of $10$\,$GM/c^{3}$. \label{fig:symmetry}}
\end{figure}

The two-dimensional flow maps reveal more interesting features. We show the $\phi$-averaged $\rho$ and $\sigma$ contours near the end of the simulations in Figure \ref{fig:rho-phi}, focusing on models $a$09$f$00, $a$09$f$01, $a$09$f$05, and $a$09$f$10. These quantities are further time-averaged to remove temporal fluctuations. We also show the corresponding $\phi$-averaged and time-averaged, radial coordinate velocity $v^{r} = u^{r}/u^{t}$ against polar angle $\theta$ in Figure \ref{fig:symmetry}, measured at $10$\,$GM/c^{2}$. We find that model $a$09$f$00 does not exhibit a jet-disk structure. The density distribution is almost spherically symmetric without any evacuated region along the poles, while the velocity streamlines are pointing towards the black hole. The radial velocity measured at $10$\,$GM/c^{2}$ is always negative and almost flat across $\theta$, indicating that the plasma close to the black hole is nearly spherically infalling. This is also consistent with the $\sigma$ contours, where we do not find strongly magnetized regions along the poles. Model $a$09$f$01 shows a jet-disk structure, but comparing to models $a$09$f$05 and $a$09$f$10, we see that the disk is thicker. Also, the bipolar outflow has a smaller opening angle, and the jets are more weakly magnetized. Even though the velocity distribution shows a anti-symmetric structure about $\theta = \pi/2$, with outflow near the poles but inflow close to the equator, the outflow velocity is lower compared to model models $a$09$f$05 and $a$09$f$10, and the velocity distribution is also flatter, suggesting a weaker and narrower jet. As the angular momentum content of the gas increases (i.e., approaching models $a$09$f$05 and $a$09$f$10), we find that highly magnetized, bipolar jets with large opening angles emerge. 


\subsection{MAD Destruction} \label{sec:disrupt}
\begin{figure*}[htb!]
    \centering
    \gridline{
    \fig{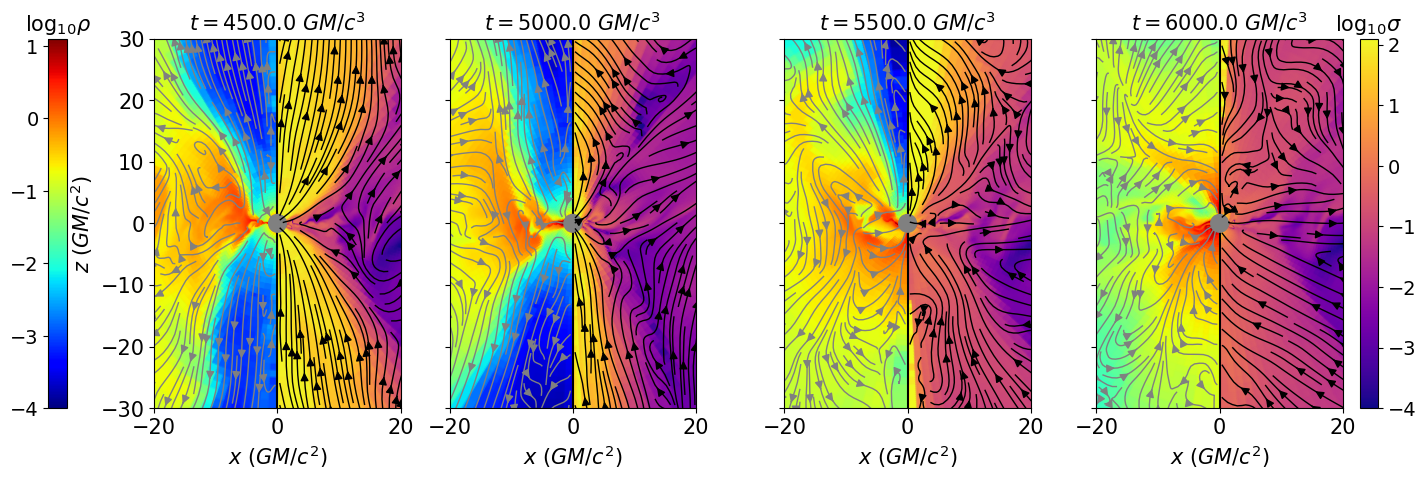}{1.0\textwidth}{(a)}}
    \gridline{
    \fig{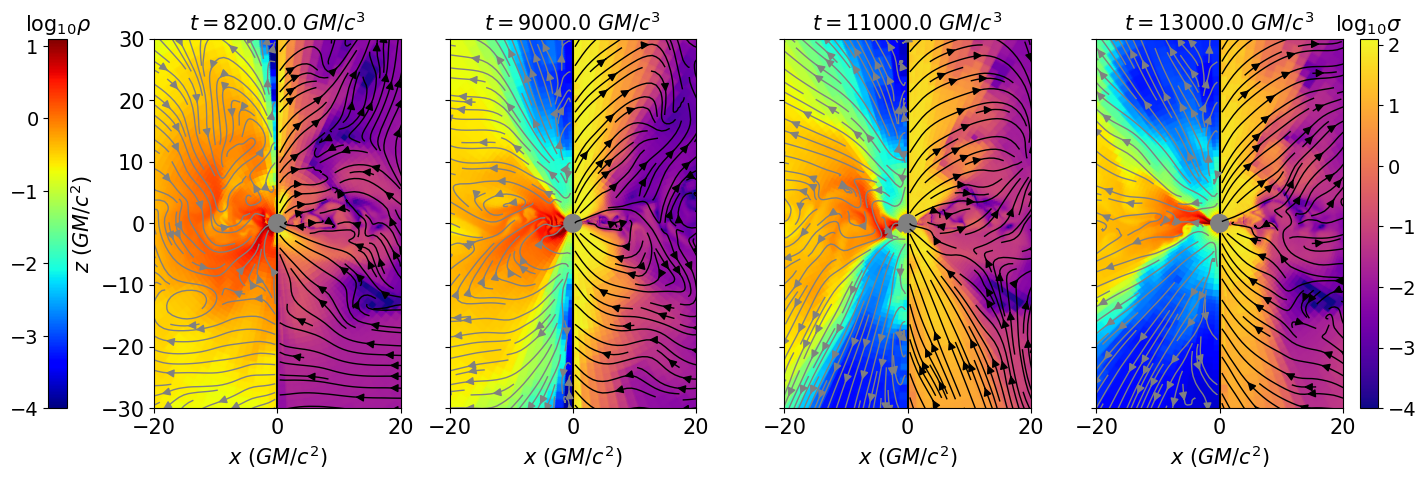}{1.0\textwidth}{(b)}}
    \caption{Snapshots of model $a$09$f$00 (a) and $a$09$f$01 (b). In each subplot, we show $\rho$ ($\sigma$)in the left (right) panel along the $x-z$ plane (log$_{10}$ scale) for the $y = 0$ slice, appended with velocity (magnetic) field lines. We show the coordinate time at the top. For model $a$09$f$00, the freely falling plasma breaks through the magnetic barrier that is set up by the horizon-saturating polodial field, loads the jet with mass, and destroys the jet-disk structure. Magnetic field lines become irregular. At the same instant, $\eta$ drops to $\sim 1$\,\% and $\phi_{\rm BH}$ drops to $\lesssim 10$. For model $a$09$f$01, the jet is reviving, and $\eta$ and $\phi_{\rm BH}$ increase from the minimum. The polar region becomes highly magnetized, evacuated, and filled with outflowing velocity streamlines. Horizon-penetrating, ordered, poloidal magnetic fields reappear.\label{fig:u3_mad_frac_0.0_spin_0.9}}
\end{figure*}

Models $a$09$f$00 and $a$09$f$01, which carry low angular momentum content, exhibit remarkable features regarding the temporal evolution of $\phi_{\rm BH}$ and $\eta$. Specifically, they do not sustain relativistic jets and do not attain the MAD state. To understand this, we show a series of $\rho$ and $\sigma$ snapshots appended with magnetic and velocity field lines for model $a$09$f$00 in Figure \ref{fig:u3_mad_frac_0.0_spin_0.9} (a). They are taken around the instances where $\phi_{\rm BH}$ and $\eta$ drop substantially. We find that the freely falling plasma breaks through the magnetic barrier set up by the horizon-saturating poloidal field and starts loading the jet with mass. This reduces plasma magnetization and weakens the jet's ability to evacuate the polar region. Magnetic field lines then become irregular, and bipolar outflow ceases to exist. 


We then turn our attention to model $a$09$f$01. This model shows a similar jet destruction pattern as $a$09$f$00: the freely falling plasma fills the polar region and loads the jet with mass. At the same time, the magnetic field lines become irregular and bipolar outflow disappears. Unlike model $a$09$f$00, however, the temporal evolution of $\phi_{\rm BH}$ and $\eta$ in $a$09$f$01 indicates a quasi-periodic cycle. To understand this, we show a series of snapshots for model $a$09$f$01 in Figure \ref{fig:u3_mad_frac_0.0_spin_0.9} (b). These snapshots are taken around the instances where $\phi_{\rm BH}$ and $\eta$ increase from their minimum. We find that the polar region, initially filled with gas, is later evacuated and populated with outflowing velocity streamlines. The plasma along the pole is initially weakly magnetized, and no significant and ordered poloidal magnetic field threads the event horizon. Afterward, the poloidal magnetic fields reappear, and the polar region becomes highly magnetized. As $\phi_{\rm BH}$ and $\eta$ increase, we observe the emergence of a jet-disk structure and the transition of magnetic field lines from an irregular to a more ordered state.


\subsection{Energy Extraction from the Black Hole} \label{sec:bzprocess}
\begin{figure}[htb!]
    \centering
    \includegraphics[width=1.0\linewidth]{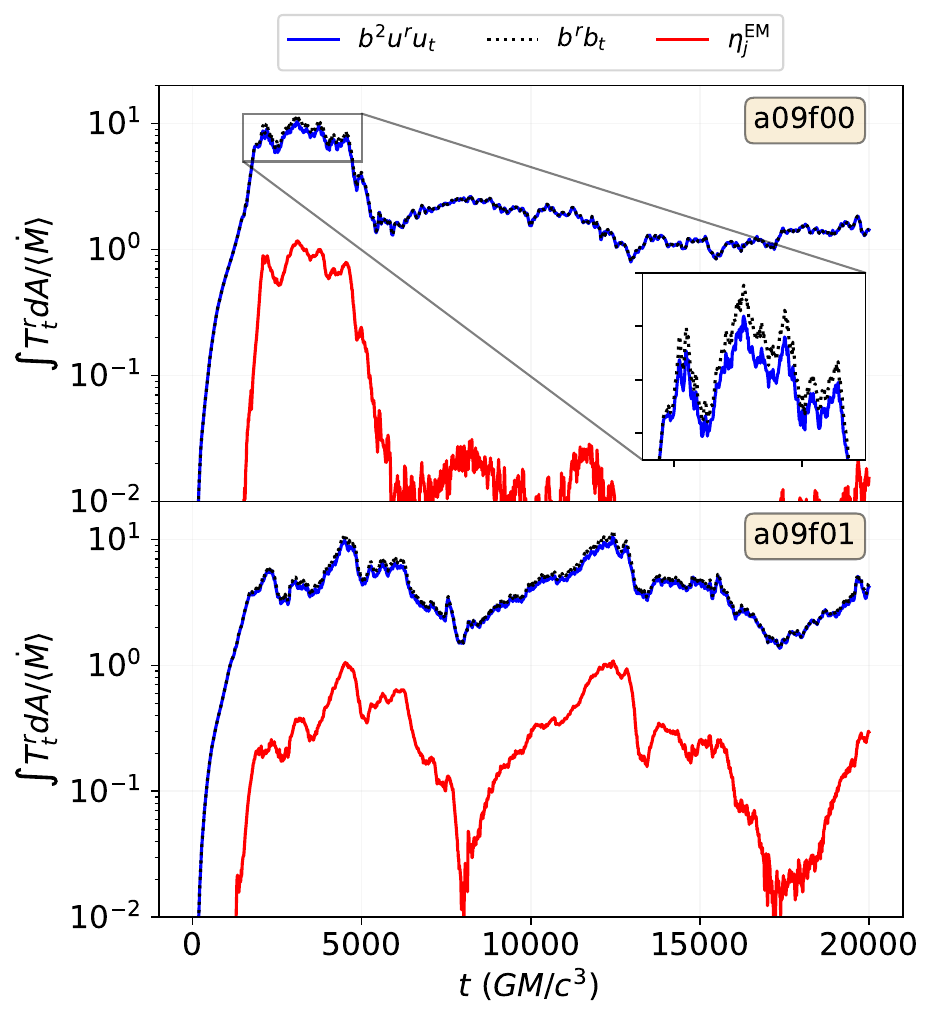}
    \caption{Black hole energy extraction efficiency $\eta^{\rm EM}_{j}$ for models $a$09$f$00 and $a$09$f$01. Here, we also decompose $(T^{r}_{t})^{\rm EM}$ into its respective contributions. When the jet is present, the $b^{r}b_{t}$ term is greater than that of $b^{2}u^{r}u_{t}$, summing to a net positive energy flux; during the jet dying phase, both terms are very close to each other and sums to a very small value. The time-variability of $\eta^{\rm EM}_{j}$ correlates with that of $\eta$ and $\phi_{\rm BH}$, in a way that $\eta^{\rm EM}_{j}$ drops significantly from $1$ when the jet is destroyed and can go negative on some occasions. \label{fig:emflux}}
\end{figure}
\begin{figure}[htb!]
    \centering
    \includegraphics[width=1.0\linewidth]{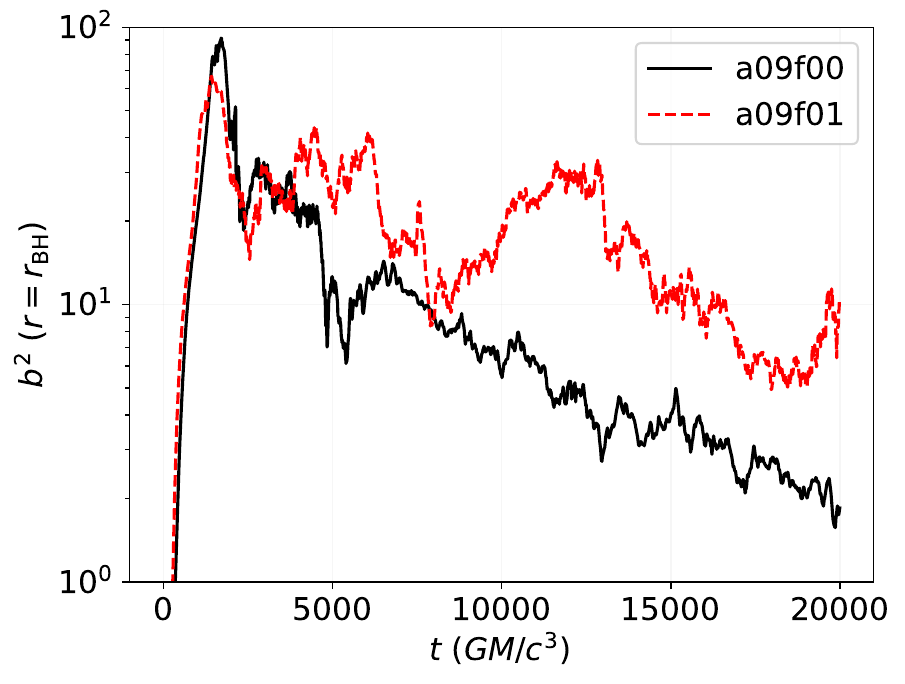}
    \caption{Radial profiles of $b^2$ evaluated at $r_{\rm BH}$ for models $a$09$f$00 and $a$09$f$01. The time-variability of the horizon magnetic energy correlates with that of $\eta^{\rm EM}_{j}$, indicating that the reduction of $\eta^{\rm EM}_{j}$ is partially due to the reduction of the magnetic field strength. \label{fig:pmageh}}
\end{figure}
\begin{figure}[htb!]
    \centering
    \includegraphics[width=1.0\linewidth]{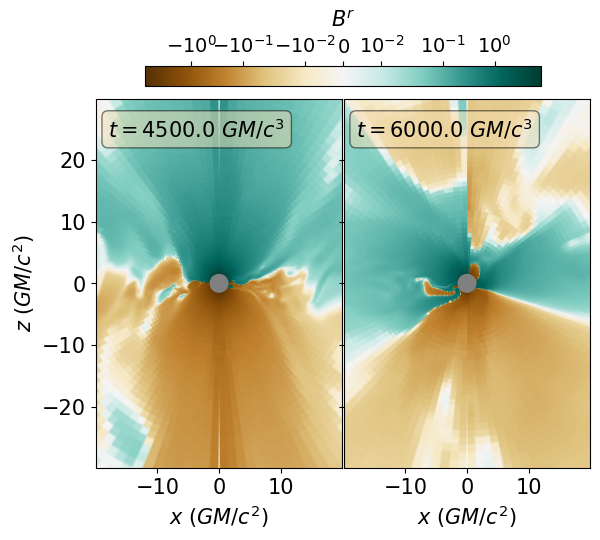}
    \caption{$B^{r}$ for model $a$09$f$00, focusing on the jet dying phase, where $\phi_{\rm BH}$ and $\eta$ decrease. The well-organized structure of the magnetic field originally presnet at $t = 4500$\,$GM/c^{3}$ is later destroyed. \label{fig:bfield-jet}}
\end{figure}
\begin{figure}[htb!]
    \centering
    \includegraphics[width=1.0\linewidth]{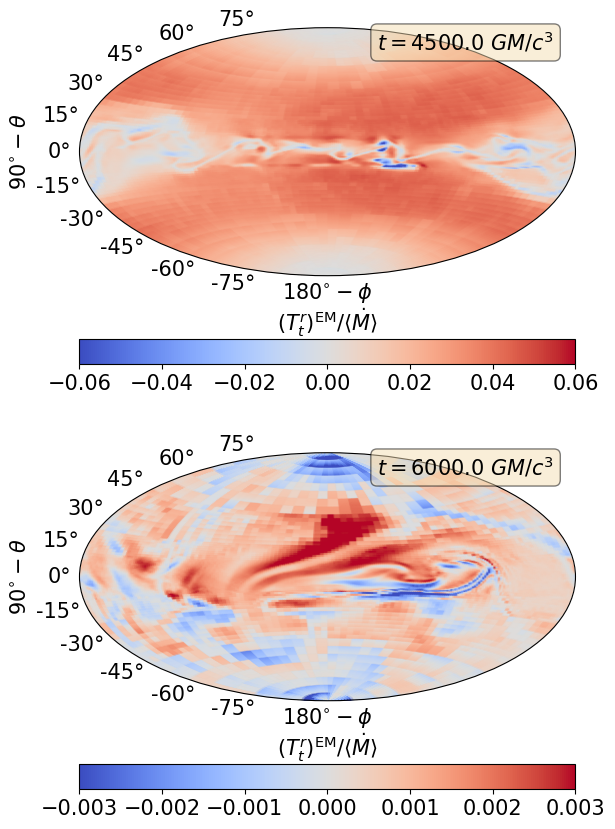}
    \caption{Distribution of the Poynting flux (normalized by $\langle\dot{M}\rangle$) evaluated at $r_{\rm BH}$ for model $a$09$f$00, focusing on the jet dying phase. As the jet disappears, the ordered, symmetric distribution turns chaotic with alternate signs, which is due to the disordered and turbulent $b^{r}$ and $u^{r}$. \label{fig:trtdA}}
\end{figure}

We showed that low angular momentum models (e.g., $a$09$f$00 and $a$09$f$01) have low $\eta$ compared to models with a higher angular momentum content (e.g., $a$09$f$07 and $a$09$f$10), hinting that they are inefficient in extracting rotation energy and angular momentum from the black hole. Here, we quantitatively examine such a efficiency. We follow \cite{dhang2024energy} to compute the jet efficiency as
\begin{equation} \label{eqn:jeteff}
    \eta^{\rm EM}_{j} = -\frac{1}{\langle \dot{M}\rangle}\int (T^{r}_{t})^{\rm EM}\sqrt{-g}d\theta d\phi, 
\end{equation}
in which $(T^{r}_{t})^{\rm EM} = b^{2}u^{r}u_{t} - b^{r}b_{t}$ is the electromagnetic part of the stress-energy tensor, and the integral is evaluated at the event horizon. The results are shown in Figure \ref{fig:emflux}. As expected, $\eta^{\rm EM}_{j}$ varies similarly to the total efficiency $\eta$ and $\phi_{\rm BH}$. For model $a$09$f$00, $\eta^{\rm EM}_{j} \sim 1$ when the jet is present, and it decreases sharply as the jet is destroyed. We note that $\eta^{\rm EM}_{j}$ can go negative, indicating electromagnetic energy flowing \textit{into} the black hole. For model $a$09$f$01, we find the same quasi-periodic time-variability, which correlates with the jet's dying and revival phases. We also decompose $(T^{r}_{t})^{\rm EM}$ into its respective contributions. When $\eta_{j}^{\rm EM}$ has its highest efficiency, the $b^{r}b_{t}$ term is larger than $b^{2}u^{r}u_{t}$, summing to a net positive energy flux, but when the jet disappear, these two terms are small and canceled each other.

There are two factors contributing to the low $\eta^{\rm EM}_{j}$ of models $a$09$f$00 and $a$09$f$01. First, the strength of the magnetic field close to the black hole is decreasing, as demonstrated by the radial profiles of $b^{2}$ measured at $r = r_{\rm BH}$ in Figure \ref{fig:pmageh}. The time-variability of the horizon magnetic energy correlates with that of $\eta^{\rm EM}_{j}$. Note that $(T^{r}_{t})^{\rm EM} \propto b^{2}$, but $\dot{M}$ does not change appreciably (see Figure \ref{fig:timeseries2}), thus explaining the drop in $\eta^{\rm EM}_{j}$. Second, there is a lack of a well-organized, structured, large-scale poloidal field threading the black hole. This is illustrated in Figure \ref{fig:bfield-jet}, focusing on model $a$09$f$00 at the instant when the jet is present ($t = 4500$\,$GM/c^{3}$) and when it disappears ($t = 6000$\,$GM/c^{3}$). We find that the well-organized structure of $B^{r}$ present at $t = 4500$\,$GM/c^{3}$ is later destroyed. The field becomes turbulent, with alternate signs of $B^{r}$ threading the horizon.

The consequences of the second factor can be understood through Figure \ref{fig:trtdA}, where the distribution of the Poynting flux at $r = r_{\rm BH}$ is shown. When the jet is present, the distribution is highly symmetric along $\theta = 90^\circ$ and is mostly positive. As the jet disappears, the magnitudes of the Poynting flux are reduced due to the weakened magnetic field. The distribution becomes irregular with alternating signs, summing to an even smaller value. We remark that the radial velocity field is also highly irregular at the horizon surface, so that the sum of the $b^{2}u^{r}u_{t}$ term is also tiny. The net result is a substantial cancellation of the $b^{2}u^{r}u_{t}$ and $b^{r}b_{t}$ terms, making the summation of the $(T^{r}_{t})_{\rm EM}$ low.


\subsection{Magnetic Flux Transport} \label{sec:fluxlost}
\begin{figure}[htb!]
    \centering
    \includegraphics[width=1.0\linewidth]{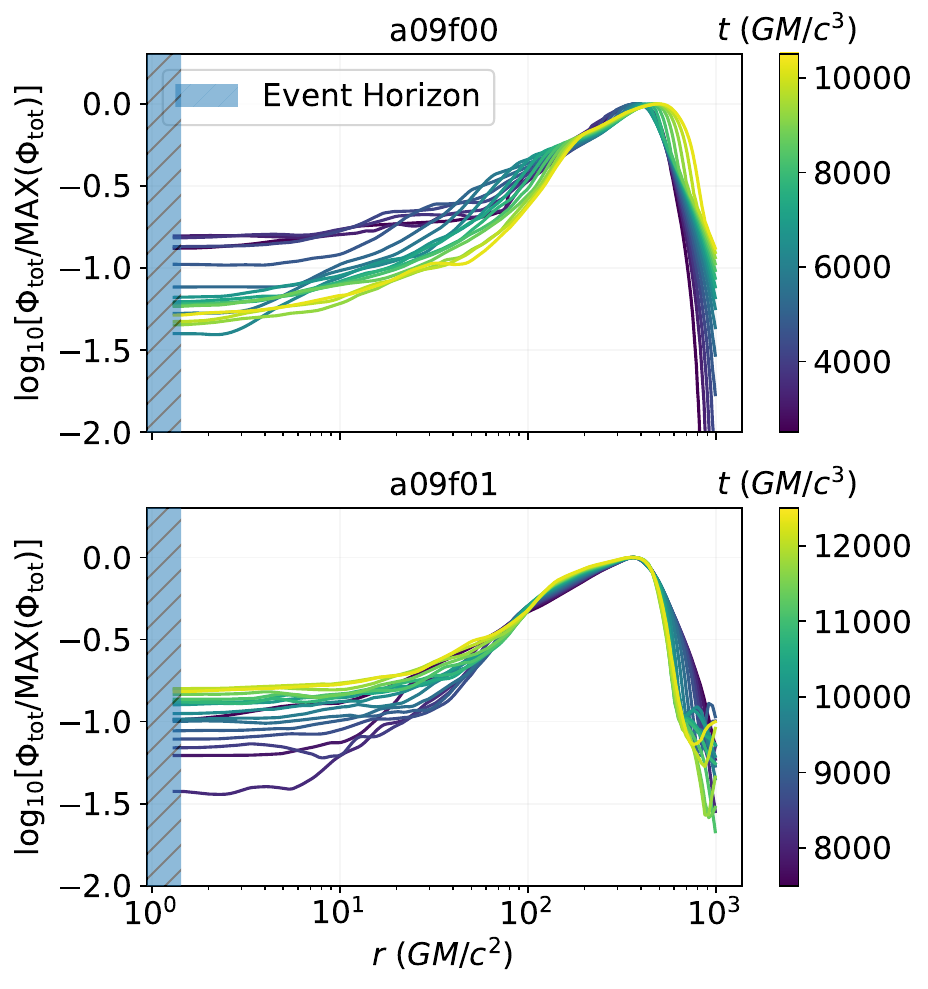}
    \caption{Time-series of radial plots of the total available magnetic flux $\Phi_{\rm tot}$ for models $a$09$f$00 and $a$09$f$01. $\Phi_{\rm tot}$ is normalized to its maximum value over all radii. For model $a$09$f$00 ($a$09$f$01), we focus on $t = 2500 - 10500$\,$GM/c^{3}$ ($t = 7500 - 12500$\,$GM/c^{3}$). The temporal evolution of $\phi_{\rm BH}$ correlates with the transport of magnetic flux. In model $a$09$f$00 ($a$09$f$01), magnetic flux is transported outward (or inward), which aligns with the drop (or increase) in $\phi_{\rm BH}$ during the same time interval. \label{fig:phitot}}
\end{figure}

We have shown that the destruction/revival of the jet-disk structure in models $a$09$f$00 and $a$09$f$01 correlates with the time variability of $\eta$ and $\phi_{\rm BH}$ and is related to the amount of poloidal magnetic field threading the event horizon. To understand how magnetic flux transport plays a role in this process, we compute the total available magnetic flux for accretion \citep{dhang2023magnetic}, $\Phi_{\rm tot}$, as
\begin{equation}
\begin{aligned}
    \Phi_{\rm tot}(r) &= \Phi_{\rm NH} + \Phi_{\rm mid}(r), \\
    \Phi_{\rm NH} &= \int_{0}^{2\pi}\int_{0}^{\pi/2}\sqrt{4\pi}B^{r}\sqrt{-g}d\theta d\phi, \\
    \Phi_{\rm mid}(r) &= \int_{r_{\rm BH}}^{r}\int_{0}^{2\pi}-\frac{\sqrt{4\pi}}{r}B^{\theta}d\phi dr,
\end{aligned}
\end{equation}
where this expression is the sum of the poloidal magnetic flux threading the northern hemisphere of the event horizon (${\rm NH}$) and the mid-plane ($\Phi_{\rm mid}$). We show the time-series of radial plots of $\Phi_{\rm tot}$ in Figure \ref{fig:phitot}.

We find that the radial plots of $\Phi_{\rm tot}$ are consistent with the temporal evolution of $\phi_{\rm BH}$. Specifically, the increase or decrease in $\phi_{\rm BH}$ correlates with the amount of $\Phi_{\rm tot}$ being brought to or escaping from the black hole. In Figure \ref{fig:phitot}(a), where we focus on the time interval of $2500 - 10500$\,$GM/c^{3}$ for model $a$09$f$00, we find that magnetic flux is transported outward, which corresponds with the significant drop in $\phi_{\rm BH}$, indicating a substantial loss of magnetic flux. After this, the magnetic flux shows no observable tendency to return, explaining why $\phi_{\rm BH}$ remains low. In Figure \ref{fig:phitot}(b), where we focus on the time interval of $7500 - 12500$\,$GM/c^{3}$ for model $a$09$f$01, we find that magnetic flux is transported inward. This is consistent with the increase in $\phi_{\rm BH}$ during the same interval, meaning the black hole acquires magnetic flux and revives the relativistic jet. The subsequent diffusion of magnetic flux is similar to Figure \ref{fig:phitot}(a) and is therefore omitted from the paper.



\subsection{Asymmetric Inflow-outflow} \label{sec:asymmetric}
\begin{figure}[htb!]
    \centering
    \includegraphics[width=1.0\linewidth]{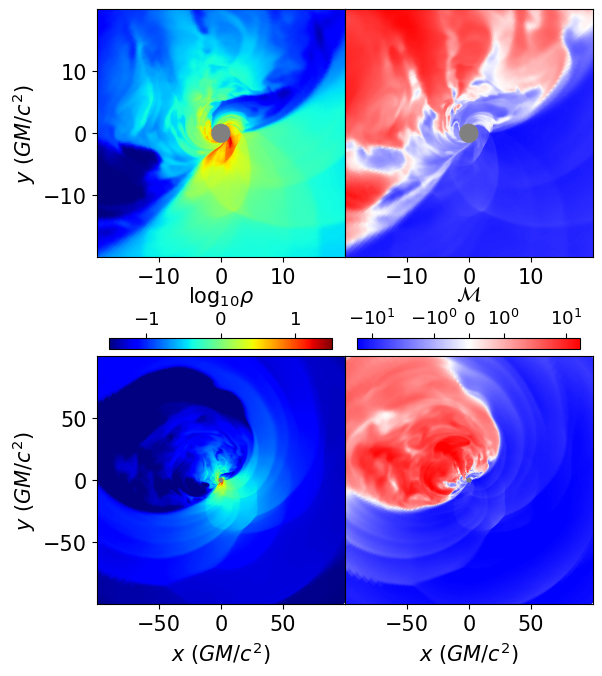}
    \caption{Snapshot in the $x-y$ plane for model $a$09$f$00 at $t = 7000$\,$GM/c^{3}$. Here, we show the density $\rho$ (log${10}$ scale) and the radial Mach number $\mathcal{M} = v^{r}/c_{s}$. The coordinate extents increase from top to bottom. Each column shares the same color scale. Low-density (also highly magnetized) cavities are ejected asymmetrically, with a speed larger than the local speed of sound. \label{fig:bubbles1}}
\end{figure}
\begin{figure*}[htb!]
    \centering
    \includegraphics[width=1.0\linewidth]{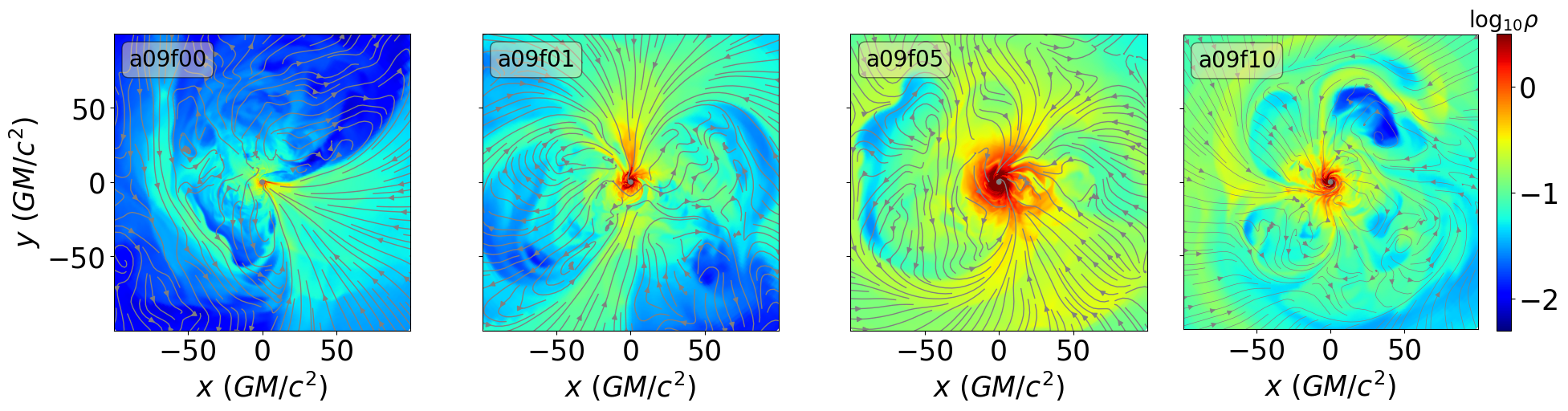}
    \caption{The density (log$_{10}$ scale) in the $x-y$ plane at $t = 18000$\,$GM/c^{3}$, appended with velocity streamlines, for models $a$09$f$00, $a$09$f$01, $a$09$f$05, and $a$09$f$10. The asymmetric inflow-outflow structure does not exist for models with larger angular momentum content of the accreting plasma. \label{fig:high-l-outflow}}
\end{figure*}

Snapshots taken along the $x-y$ plane reveal how models with low angular momentum content lose poloidal magnetic flux. We show the density and radial Mach number $\mathcal{M} = v^{r}/c_{s}$ for model $a$09$f$00 in Figure \ref{fig:bubbles1}. The snapshot is taken at $t = 7000$\,$GM/c^{3}$, which is shortly after a huge amount of magnetic flux is lost. We find gigantic, asymmetrical cavities of low-density gas being expelled from the black hole in the second quadrant. The low density of the cavities implies that they are also highly magnetized and thus carry a substantial amount of magnetic flux. The $\mathcal{M}$ contours show a highly asymmetric inflow-outflow structure — plasma converges to the black hole for $\phi \sim -135^\circ$ to $45^\circ$ but is expelled at $\phi \sim 45^\circ$ to $225^\circ$. Close to the black hole, the expulsion speed is larger than the local speed of sound.

Figure \ref{fig:bubbles1} thus reveals the reason for the sudden magnetic flux loss at $\sim 5000$\,$GM/c^{3}$ and the low $\phi_{\rm BH}$ afterward, specifically for model $a$09$f$00. At the beginning of the simulation, the free-falling plasma brings huge amounts of magnetic flux to the black hole in a very short time. The magnetic flux quickly saturates and generates an outward, asymmetric `bounce' that pushes the mass and magnetic flux away from the black hole. After the asymmetric inflow-outflow structure is established, the inflow of plasma and magnetic flux approximately balances the expulsion, making both $\dot{M}$ and $\phi_{\rm BH}$ remain roughly constant. Since the expelled bubbles are much less dense than their surroundings (with a density contrast of $\sim 10^{-1}$ to $10^{-2}$), they are prone to buoyancy and cannot easily return to the black hole.

Figure \ref{fig:high-l-outflow} shows the density in the $x-y$ plane with velocity streamlines, focusing on models $a$09$f$00, $a$09$f$01, $a$09$f$05, and $a$09$f$10. It is taken at $t = 18000$\,$GM/c^{3}$. The asymmetric inflow-outflow structures exist for model $a$09$f$01 but are not seen  in models $a$09$f$05 and $a$09$f$10. Thus, models with a sufficient amount of angular momentum do not lose flux as much as, and as easily as, models $a$09$f$00 and $a$09$f$01. The accretion dynamics thus approaches that of the standard FM torus, where a sustainable relativistic jet is formed and the accretion can attain the MAD state. 

We remark that the development of the asymmetric inflow-outflow structure is probably related to the angular momentum content of the plasma. The angular momentum provides a centrifugal barrier that prevents the prompt infall of magnetic field-carrying gas to the black hole. The black hole can then slowly saturate with magnetic flux rather than being overly loaded with magnetic field and violently blasting plasma outward. We also note that this structure is responsible for keeping $\dot{M}$ low even when the flow is not magnetically arrested.



\subsection{Magnetic Bubbles and the Role of Shear} \label{sec:bubbles}

Accretion in the MAD state is typically  characterized by regular flux eruptions \citep{white2019resolution,dexter2020sgr,porth2021flares,begelman2022really,scepi2024magnetic}, where magnetic flux-carrying cavities (or bubbles) are expelled from the black hole. The ejected flux slowly returns through the main accretion flow and saturates the black hole again. However, our low angular momentum models demonstrate that the expelled magnetic flux has difficulty returning to the black hole.

\begin{figure*}[htb!]
    \centering
    \gridline{
    \fig{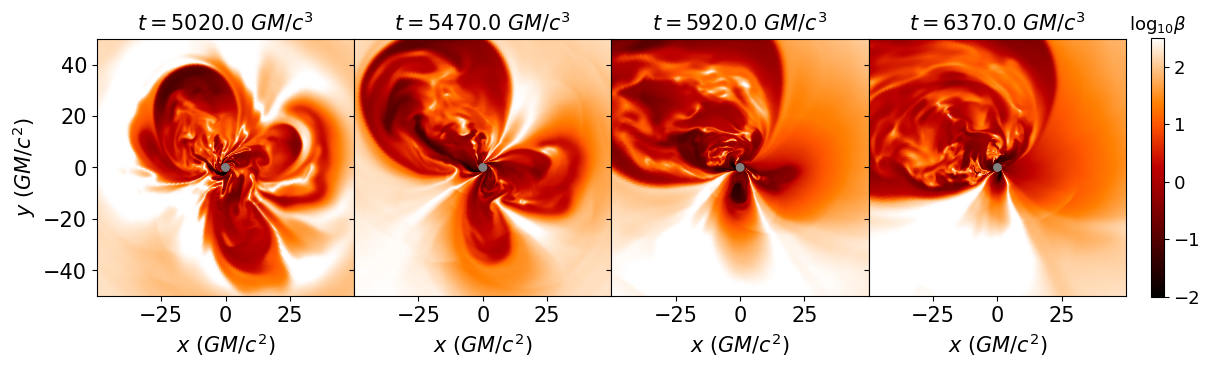}{1.0\textwidth}{(a)}}
    \gridline{
    \fig{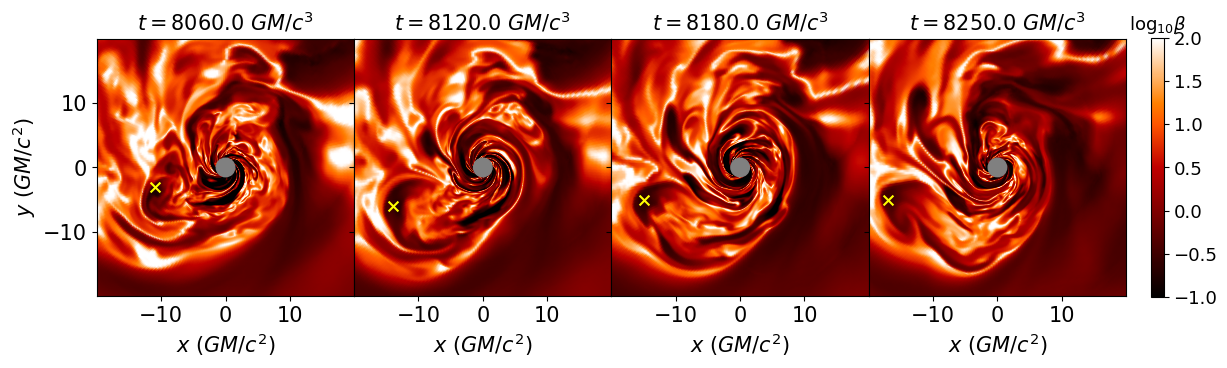}{1.0\textwidth}{(b)}}
    \caption{Snapshots of the plasma $\beta$ in the $x-y$ plane. We show (a) model $a$09$f$00 and (b) model $a$09$f$05. The coordinate time is labeled at the top of each subplot. For (a), magnetic flux is lost via gigantic, asymmetrical magnetic bubbles. After the bubbles are ejected in a preferred direction, they show no tendency to return to the black hole. Some initially ejected bubbles are caught up by inflowing streamlines but are immediately ejected again, so the horizon-penetrating flux remains small. For (b), we zoom into a domain of $[(-20, 20)\times(-20,20)]$. Once the bubbles are ejected, they are elongated by shear, circularize around the hole, and decrease in magnetic dominance. Magnetic flux is mixed into the surrounding inflowing plasma and returns to the hole. We label an example of such a bubble with a yellow cross marker. \label{fig:bubbles_a09f00}}
\end{figure*}

To understand why, we track the motion of the bubbles via a series of snapshots of the plasma $\beta$ along the $x-y$ plane in Figure \ref{fig:bubbles_a09f00} (a), focusing on model $a$09$f$00. We concentrate on the instant when the gigantic, asymmetrical bubbles are launched, and magnetic flux is substantially lost. After being ejected from the black hole, the magnetized bubbles show no tendency to circularize and/or stop expanding, and they quickly exit the domain of interest. We also find that some bubbles ejected before $t = 5020$\,$GM/c^{3}$ are caught up by the inflowing streamlines. However, once they return to the black hole, they are immediately expelled. These snapshots thus demonstrate how magnetic flux is being transported outward for models with low angular momentum content. The asymmetric outflow structure is responsible for expelling the magnetic flux available for accretion. As such, it is difficult for the horizon-penetrating magnetic flux to reach a high value.

We next turn our attention to a model with more angular momentum content, $a$09$f$05, where we also show a series of plasma $\beta$ snapshots in Figure \ref{fig:bubbles_a09f00} (b). Unlike model $a$09$f$00, we do not find any large-scale bubbles floating away from the black hole. We also track magnetic bubbles generated via flux eruptions. Compared to model $a$09$f$00, however, their spatial scale is much smaller. Once these bubbles are ejected, they are azimuthally torn by the velocity shear of the gas, while the magnetic dominance of the bubbles is reduced substantially. The bubbles are then mixed with the surrounding inflowing plasma and are slowly recycled and absorbed by the black hole.

These snapshots demonstrate the qualitative difference between the low and high angular momentum models in terms of the motion of the bubbles and the recycling of magnetic flux. Velocity shear is likely an important parameter in this case. We argue that the azimuthal velocity shear could easily elongate the bubbles, reducing their magnetic dominance. Additionally, shear promotes mixing between weakly and highly magnetized plasma, effectively redistributing the magnetic flux in space -- it stirs the expelled flux into the surrounding weakly magnetized plasma that follows the inflowing streamlines. Without velocity shear, the bubbles would be expelled radially and float away to larger radii due to buoyancy.


\section{Discussion} \label{sec:physical}
%


Our work, which is novel in its use of a torus with variable angular velocity as the initial condition to survey accretion onto spinning black holes with different angular momentum contents, shows results that are broadly consistent with previous studies (e.g., \citet{ressler2021magnetically} and \citet{2023arXiv231011487L}). Specifically, we show that low angular momentum accretion struggles to sustain relativistic jets. The model with the lowest amount of angular momentum ($a$09$f$00) has an $\eta$ of roughly $100$\% and launches relativistic jets, but only for a very short duration. Then, the free-falling plasma breaks through the magnetic barrier, loads the jet with mass, and destroys the jet. The horizon gas outflow efficiency remains at a few percent or less. The model with a small but non-zero amount of angular momentum ($a$09$f$01) shows quasi-periodic oscillations in the temporal evolution of $\eta$, which manifest as the episodic revival and decay of bipolar outflows/jets in the snapshots. Additionally, low angular momentum accretion is unable to attain the MAD state. Model $a$09$f$00 attains a saturated $\phi_{\rm BH} \sim 50$, but then it drops substantially (along with the reduction $\eta$) to a level of $\lesssim 10$. Model $a$09$f$01 shows a quasi-periodic cycle of $\phi_{\rm BH}$, oscillating between $\phi_{\rm BH} \lesssim 10$ and $50$, and it is highly correlated with the time-variability of the horizon gas outflow efficiency $\eta$. Thus, the assertion that accreted angular momentum content of the plasma is a crucial parameter for sustaining jets and attaining the MAD state should be quite general across different initial conditions. 

Despite repeating major results of previous works, we also have some new findings. Specifically, we find that the `lifetime' of the saturation state of $\phi_{\rm BH} \sim 50$ is shorter compared to the works mentioned above. Model $a$09$f$00 has a saturated $\phi_{\rm BH} \sim 50$ only for a few thousand $GM/c^3$, while $\phi_{\rm BH}$ for model $a$09$f$01 shows a quasi-periodic cycle, so the time of $\phi_{\rm BH}$ staying at roughly $50$ will be even shorter. This is in contrast to the work by \citet{2023arXiv231011487L}, where $\phi_{\rm BH}$ saturates at $\sim 50$ for a few ten-thousands $GM/c^{3}$. Also, low angular momentum accretion models do not maintain a jet-disk structure. For model $a$09$f$00, as $\phi_{\rm BH}$ and $\eta$ drop, the jet-disk structure completely disappears, and that plasma is radially in-falling to the black hole within the region of our interest. For model $a$09$f$01, the jet-disk structure exhibits a quasi-periodic cycle of revival and destruction, and it is highly correlated with the time-variability in $\phi_{\rm BH}$ and $\eta$. Comparing time-series and snapshots plots also reveal that the accretion dynamics starts to approach that of a standard FM torus when $f$ increase towards $1$. Our work thus qualitatively shows how angular momentum content governs accretion dynamics and the launching of relativistic jets.

These discrepancies could be due to the different initial conditions assumed, which include factors such as the density distribution, the configuration and strength of the magnetic field, and the initial magnetic flux distribution. For instance, given that the ratio of the radius of the pressure maximum to its `Bondi radius' is less than $1$, it suggests that the timescale over which most of the gas begins to plunge into the black hole significantly differs from previous studies, where this ratio is typically greater than $1$. Additionally, we note that our initial magnetic field is relatively weak, which makes it easier for the gas to penetrate the magnetic barrier. Additionally, previous studies adopted a uniform vertical magnetic field extending across the entire domain as the initial condition, requiring plasma to cross magnetic field lines to populate the pole. In our study, however, the initial magnetic field is a loop entirely contained within the torus. Furthermore, if a significant amount of magnetic flux exists at a radius closer to the black hole than where most of the gas resides, the magnetic field strength can be amplified via flux-freezing, effectively blocking the free-falling plasma. Due to the high computational cost of 3D GRMHD simulations, we could only explore a limited range of parameter space. As a result, we are unable to identify the primary governing parameters responsible for the observed differences. Future studies should investigate how zero angular momentum accretion is influenced by the factors outlined here. 

Our results, or in general, the study of low angular momentum accretion onto black holes, could help us interpret observational results from some astrophysical sources. In particular, the absence of radio jets from Sgr~A* is still puzzling. Comparing time-variability data to simulations suggests that the accretion is possibly wind-fed by Wolf-Rayet stars \citep{murchikova2022remarkable}, where the plasma contains a small amount of angular momentum \citep{2018MNRAS.478.3544R, 2020MNRAS.492.3272R}. The absence of a detectable relativistic jet could be explained by its dissipation via the kink instability \citep{ressler2021magnetically} within a few hundred gravitational radii. We suggest here that another possible explanation is that the relativistic jet is either never launched (model $a$09$f$00) or launched for a very short duration (model $a$09$f$01). Given that the current EHT images of Sgr~A* cover only the near-horizon region where the frame-dragging effects and plunging orbits dominate the accretion flow \citep{2022ApJ...930L..15E}, it would be difficult to rule out either the no-jet or the kink dissipation model. A future space-based VLBI mission, such as the Black Hole Explorer project\citep{johnson2024black}, could help determine whether Sgr~A* does contain a relativistic jet. Additionally, it would be interesting to test whether low angular momentum accretion onto spinning black holes can solve the over-variability problem of the $230$\,GHz light curves predicted by theoretical models of Sgr~A* \citep{2022ApJ...930L..16E}, as an alternative model to the changing of electron temperature \citep{chan2023230, chan2024230}.

\section{Summary} \label{sec:conclude}

To examine how the accreted angular momentum content governs the accretion dynamics of strongly magnetized plasma, we performed three-dimensional GRMHD simulations of accreting, rapidly spinning black hole ($a = +0.9$), in which the initial condition is an FM torus with the angular velocity being a fraction $f$ of the standard case, and the torus is threaded with a large amount of magnetic flux. Our simulations span $f = 0.0$ to $f = 1.0$, covering plasma with low, intermediate, and high angular momentum content. We aim to determine the reason for the inefficiency of these models in maintaining the MAD state and relativistic jets. Our major findings are:

\begin{enumerate}

\item \textbf{Low angular momentum accretion models are inefficient in extracting rotational energy from the black hole to drive relativistic jets}. The electromagnetic Poynting flux reduces in magnitude due to reduction of the horizon magnetic field strength. Also, the horizon magnetic and velocity field becomes irregular and turbulent, causing $u^{r}$ and $b^{r}$ to change sign across the horizon. This makes the surface integral of $(T^{r}_{t})_{\rm EM}$ tiny, thereby reducing the \textit{net} energy flowing out from the black hole. 

\item \textbf{The time-variability of $\phi_{\rm BH}$ and $\eta$ relates to the inward/outward transport of magnetic flux}. Time-series plots of the total available magnetic flux for accretion $\Phi_{\rm tot}$ indicates that as $\phi_{\rm BH}$ and $\eta$ drop (increase), magnetic flux is being transported outward (inward). As such, the reason that low angular momentum models are weak at attaining the MAD state and maintaining relativistic jets is because of the loss of horizon-penetrating magnetic fields.

\item \textbf{The asymmetric inflow-outflow structure is responsible for the loss of magnetic flux}. An asymmetric inflow-outflow structure develops for low angular momentum models. A huge amount of magnetic flux is expelled in a preferred direction via gigantic, asymmetrical bubbles with high speed. Subsequently, magnetic flux finds it difficult to accumulate on the black hole. Such a structure ceases to exist when the angular momentum content of the plasma is higher.

\item \textbf{Velocity shear is important in governing the magnetic flux recycle}. Magnetic bubbles generated by flux eruptions in low angular momentum models will escape from the black hole quickly and freely due to being prone to buoyancy, while those generated in the high angular momentum models will be torn azimuthally by velocity shear, reducing magnetic dominance, and circularize around the black hole. Velocity shear also promotes mixing between weakly and strongly magnetized gas, stirring magnetic fields into surrounding magnetized plasma that follows the main accretion flow. 

\end{enumerate}

Future work, including polarized raytracing GRMHD snapshots and spectral modeling, could provide more information on the dependence of the electromagnetic observables on the angular momentum contents of black hole accretion flows.

\begin{acknowledgments}
Ho-Sang (Leon) Chan acknowledges support from the Croucher Scholarship for Doctoral Studies by the Croucher Foundation. MB and JD acknowledge support from NASA Astrophysics Theory Program grants 80NSSC22K0826 and 80NSSC24K1094.
\end{acknowledgments}


\bibliography{main}{}
\bibliographystyle{aasjournal}


\end{document}